\documentclass[journal]{IEEEtran}

\usepackage{siunitx}								% for pdf, bitmapped graphics files
\usepackage{epsfig} 									% for postscript graphics files
\usepackage{mathptmx} 		% assumes new font selection scheme installed
\usepackage{times} 									% assumes new font selection scheme installed
\usepackage{hyperref}
\usepackage{amsmath} 								% assumes amsmath package installed
\usepackage{amssymb,amsfonts} 					% assumes amsmath package installed
\usepackage{flushend} 								%to balance columns on each page(especially the last page)
\usepackage{cite}
\usepackage{placeins}
\usepackage{graphicx}
\usepackage{color,soul}
\usepackage{graphicx}
\usepackage{algorithm2e}
\usepackage{color}
\usepackage{verbatim}
\usepackage{array}
\usepackage{wrapfig}
\usepackage{flushend}
\usepackage{subfigure}
\usepackage{paralist}
\usepackage{mathtools}
\usepackage{multirow}

\newcommand{\specialcell}[2][c]{%
 \begin{tabular}[#1]{@{}c@{}}#2\end{tabular}}
\begin{document}
\title{The COMMOTIONS Urban Interactions Driving Simulator Study Dataset}
\author{Aravinda Ramakrishnan Srinivasan$^{1*}$, Julian Schumann$^{2}$, Yueyang Wang$^{1}$, Yi-Shin Lin$^{1}$, Michael Daly$^{1}$, Albert Solernou$^{1}$, Arkady Zgonnikov$^{2}$,  Matteo Leonetti$^{3}$, Jac Billington$^{4}$, and Gustav Markkula$^{1*}$% 
\thanks{This project has received funding from UK Engineering and Physical Sciences Research Council under fellowship named COMMOTIONS - \\Computational Models of Traffic Interactions for Testing of Automated Vehicles - EP/S005056/1.}
\thanks{This research was partially supported by TAILOR, a project funded by EU Horizon 2020 research and innovation programme under GA No 952215. More particularly, Julian Schumann was supported by the TAILOR connectivity fund.}
\thanks{For the purpose of open access, the author(s) has applied a Creative Commons Attribution (CC BY) license to any Accepted Manuscript version arising.}% 
\thanks{$^{1}$ Institute for Transport Studies, University of Leeds, UK}%
\thanks{$^{2}$ Department of Cognitive Robotics, Delft University of Technology, Netherlands}%
\thanks{$^{3}$ Department of Informatics, King's College London, UK}%
\thanks{$^{4}$ School of Psychology, University of Leeds, UK}%
\thanks{\href{https://osf.io/eazg5}{Open Science Foundation - Data repository link}}
\thanks{$^{*}$ Corresponding authors: {\tt\small A.R.Srinivasan@leeds.ac.uk, G.Markkula@leeds.ac.uk}}%
}

% The paper headers
\markboth{Open Science Foundation Dataset Paper}{Srinivasan \MakeLowercase{\textit{et al.}}: The COMMOTIONS Urban Interactions Driving Simulator Study Dataset}
\maketitle
\begin{abstract}
Accurate modelling of road user interaction has received lot of attention in recent years due to the advent of increasingly automated vehicles. To support such modelling, there is a need to complement naturalistic datasets of road user interaction with targeted, controlled study data. This paper describes a dataset collected in a simulator study conducted in the project COMMOTIONS, addressing urban driving interactions, in a state of the art moving base driving simulator. The study focused on two types of near-crash situations that can arise in urban driving interactions, and also collected data on human driver gap acceptance across a range of controlled gap sequences.  
\end{abstract}
\section{Introduction}
Public roadways are slowly seeing introduction of vehicles with advanced automation recently~\cite{noauthor_mercedes-benz_2023}. Vehicles with different automation capabilities~\cite{noauthor_sae_nodate} have to interact with each other and other vulnerable road users (VRU) to ensure safety and comfort for everyone involved. In order to interact efficiently the automation needs to understand the human road user interactive behaviours. Accurate models of human road user behaviours can help car manufacturers to incorporate human-like behaviours into their driver assistance and automation systems. It is also an useful tool for regulators to properly safety test new driving technology at software level before real life deployment of automation technologies. 

These models of human interaction need to capture not just regular driving interactions, but also human behaviour in situations where the interactions deteriorate into near-crash situations. In existing naturalistic driving datasets on driving interactions, e.g., ~\cite{inDdataset,rounDdataset,highDdataset,noauthor_traffic_nodate,zhan2019constructing} there is a paucity of near-crash data. In contrast, there is a long tradition of studying near-crash driving interactions in driving simulators, e.g., \cite{mcgeheeetal1999examination,huli2017drivers,lietal2019how}, but these datasets have typically not been made publicly available. This paper describes a dataset collected in a high-fidelity driving simulator as part of the UK EPSRC COMMOTIONS project, with the primary aim of permitting testing of human driving interaction models on near-crash human behaviour data. The goal was to collect a relatively large number of observations, for two specific targeted safety-critical interactions, each at two different levels of urgency, since urgency is known to be an important factor in human near-crash response \cite{markkulaetal2016farewell}. A secondary aim of the study was to collect data on human driver gap acceptance in vehicle flows with controlled gap sequences. 

The aim of this paper is simply to describe the study and the dataset; analysis and modelling of the collected data is being pursued for conventional publication elsewhere. Below, we describe the simulated scenarios, the driving simulator, the participants, the experimental procedure, as well as how to access and use the data.

\section{Simulated scenarios}
\begin{table*}    
 \caption{Gaps between two consecutive vehicles on the tangential road used in the gap acceptance scenario, given in \SI{}{m}.}
 \centering
 \begin{tabular}{c c c c c c c c c c c c c c c c c} 
 \hline
 \specialcell{Variation/Gaps} & 1 & 2 & 3 & 4 & 5 & 6 & 7 & 8 & 9 & 10 & 11 & 12 & 13 & 14 & 15 & 16  \\ [0.2ex] 
 \hline
1  & 93 & 70 & 27.5 & 87.1 & 25 & 73.7 & 25 & 25 & 25 & 25 & 73.7 & 113.9 & 25 & 25 & 25 & 25 \\
2  & 93 & 70 & 27.5 & 78.2 & 25 & 64.8 & 25 & 25 & 25 & 25 & 64.8 & 113.9 & 25 & 25 & 25 & 25 \\
3  & 93 & 70 & 27.5 & 69.2 & 25 & 55.8 & 25 & 25 & 25 & 25 & 55.8 & 113.9 & 25 & 25 & 25 & 25 \\
4  & 93 & 70 & 27.5 & 69.2 & 25 & 46.9 & 25 & 25 & 25 & 25 & 46.9 & 113.9 & 25 & 25 & 25 & 25 \\
5  & 93 & 70 & 27.5 & 60.3 & 25 & 73.7 & 25 & 25 & 25 & 25 & 73.7 & 113.9 & 25 & 25 & 25 & 25 \\
6  & 93 & 70 & 27.5 & 60.3 & 25 & 64.8 & 25 & 25 & 25 & 25 & 64.8 & 113.9 & 25 & 25 & 25 & 25 \\
7  & 93 & 70 & 27.5 & 60.3 & 25 & 55.8 & 25 & 25 & 25 & 25 & 55.8 & 113.9 & 25 & 25 & 25 & 25 \\
8  & 93 & 70 & 27.5 & 60.3 & 25 & 46.9 & 25 & 25 & 25 & 25 & 46.9 & 113.9 & 25 & 25 & 25 & 25 \\
9  & 93 & 70 & 27.5 & 51.4 & 25 & 73.7 & 25 & 25 & 25 & 25 & 73.7 & 113.9 & 25 & 25 & 25 & 25 \\
10 & 93 & 70 & 27.5 & 51.4 & 25 & 64.8 & 25 & 25 & 25 & 25 & 64.8 & 113.9 & 25 & 25 & 25 & 25 \\
11 & 93 & 70 & 27.5 & 51.4 & 25 & 55.8 & 25 & 25 & 25 & 25 & 55.8 & 113.9 & 25 & 25 & 25 & 25 \\
12 & 93 & 70 & 27.5 & 51.4 & 25 & 46.9 & 25 & 25 & 25 & 25 & 46.9 & 113.9 & 25 & 25 & 25 & 25 \\
13 & 93 & 70 & 27.5 & 42.4 & 25 & 73.7 & 25 & 25 & 25 & 25 & 73.7 & 113.9 & 25 & 25 & 25 & 25 \\
14 & 93 & 70 & 27.5 & 42.4 & 25 & 64.8 & 25 & 25 & 25 & 25 & 64.8 & 113.9 & 25 & 25 & 25 & 25 \\
15 & 93 & 70 & 27.5 & 42.4 & 25 & 55.8 & 25 & 25 & 25 & 25 & 55.8 & 113.9 & 25 & 25 & 25 & 25 \\
16 & 93 & 70 & 27.5 & 42.4 & 25 & 46.9 & 25 & 25 & 25 & 25 & 46.9 & 113.9 & 25 & 25 & 25 & 25 \\

 \hline
 \label{tab:gapTable}
 \end{tabular}
 % \vspace{-0.6cm}
\end{table*}
The simulated urban driving consisted of several routine and two near-crash scenarios, each with two urgency variations. For each type of near-crash scenario, there was also a non-critical version of the same scenario, to which the drivers were exposed as part of the non-critical driving, before experiencing the safety-critical scenarios.

There was one subject vehicle (SV) controlled and driven by the participant. There were several vehicles which were programmed with pre-determined motion in the routine scenarios and one principal other vehicle (POV) in each of the near-crash and near-crash like routine scenarios. The scenarios happened in an urban-like simulated environment with a speed limit of $30$ miles/hour ($13.41$ m/s). The speed limit is the prevalent urban speed limit~\cite{noauthor_speed_nodate} in the United Kingdom (UK). The participants were given route directions via a GPS-like pre-programmed voice prompt.

\subsection{Routine scenarios}
\subsubsection{Gap acceptance by SV}\label{sec:gap_accept}

\begin{figure}
    \centering\includegraphics[scale = 0.4]{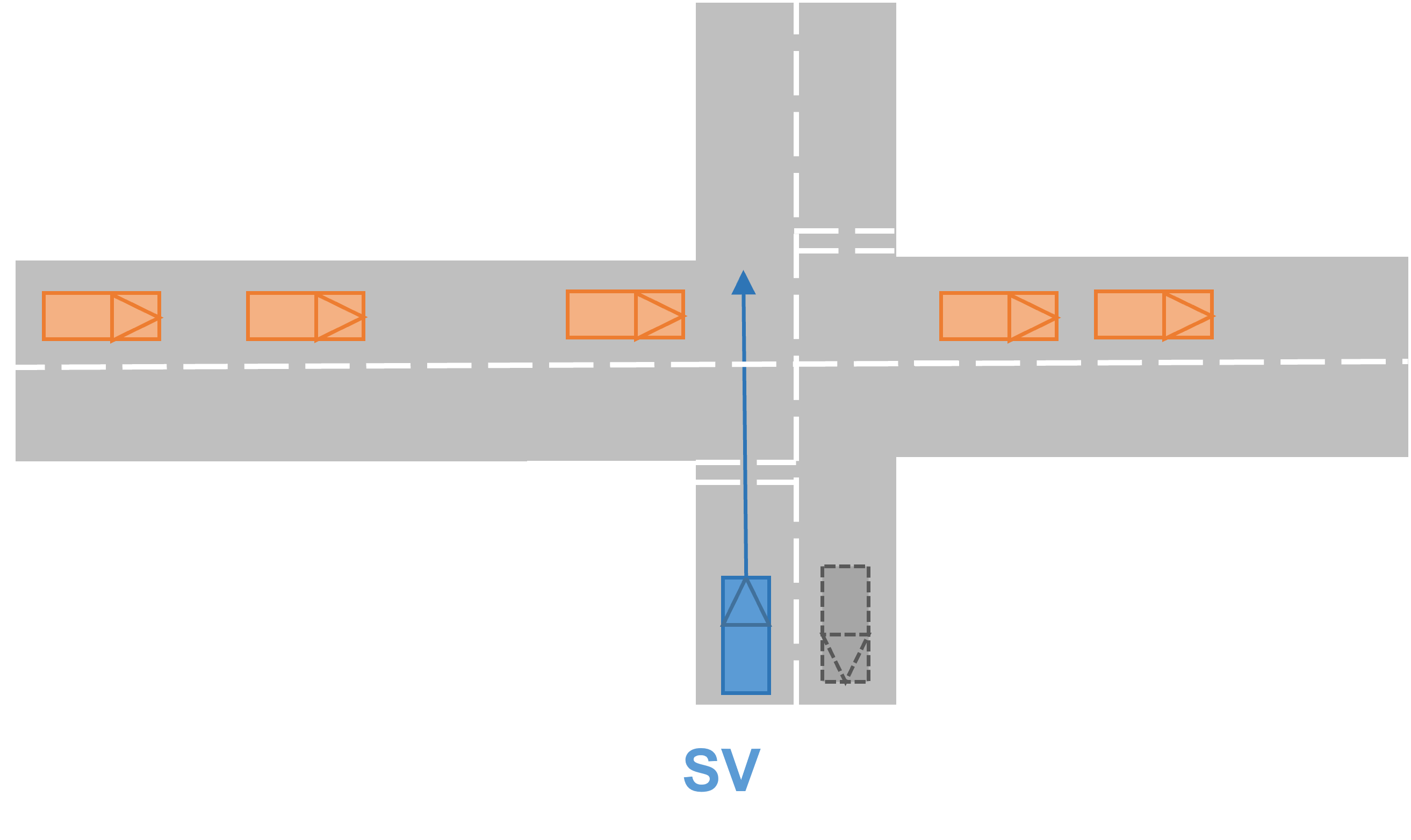}
    \caption{A graphical representation of the gap acceptance scenario where SV has no priority and POVs are in the arterial (tangential) roadway with priority}
    \label{fig:gap_accept}
\end{figure}

One routine scenario consisted of SV passing through a crossroad without priority. The SV has to accept one of the gap presented by a stream of vehicles (driving at a constant velocity of \SI{13.4}{m/s} in the arterial (tangential) road. This scenario recreates a typical gap acceptance situation as graphical described in \figurename~\ref{fig:gap_accept}. The available gaps were fixed and there were a total of sixteen variations repeated across participants (between subject variation). The gaps between the vehicles are presented in~\tablename~\ref{tab:gapTable}, where the vehicles are triggered in such a way that with a constant velocity the SV would reach the intersection when gap 3 just opened. The gap 1 is between the intersection and the first vehicle to cross the intersection in the tangential road. 

\subsubsection{Other routine scenarios}\label{sec:other_sec}
The SV was shown other routine scenarios like pulling into a signalised crossroads where the signal was initially red for the SV lane. As the SV approached the junction, the traffic lights changed to green for the SV lane allowing the SV to pass through without having to come to a full stop. Another routine scenario was the exact opposite wherein the signal turned from green to red as the SV approached the junction. This would make the SV come to a full stop like in a typical urban roadways. Finally, there was a three way intersection where the SV was instructed through the GPS like voice prompt to turn left. 
\subsection{Near-crash like routine scenarios}
\subsubsection{Yielding by POV to SV turning against traffic in a T-like junction}\label{sec:like_yield}
\begin{figure}
    \centering
    \subfigure{
    \includegraphics[scale =0.4, width = 0.3\linewidth]{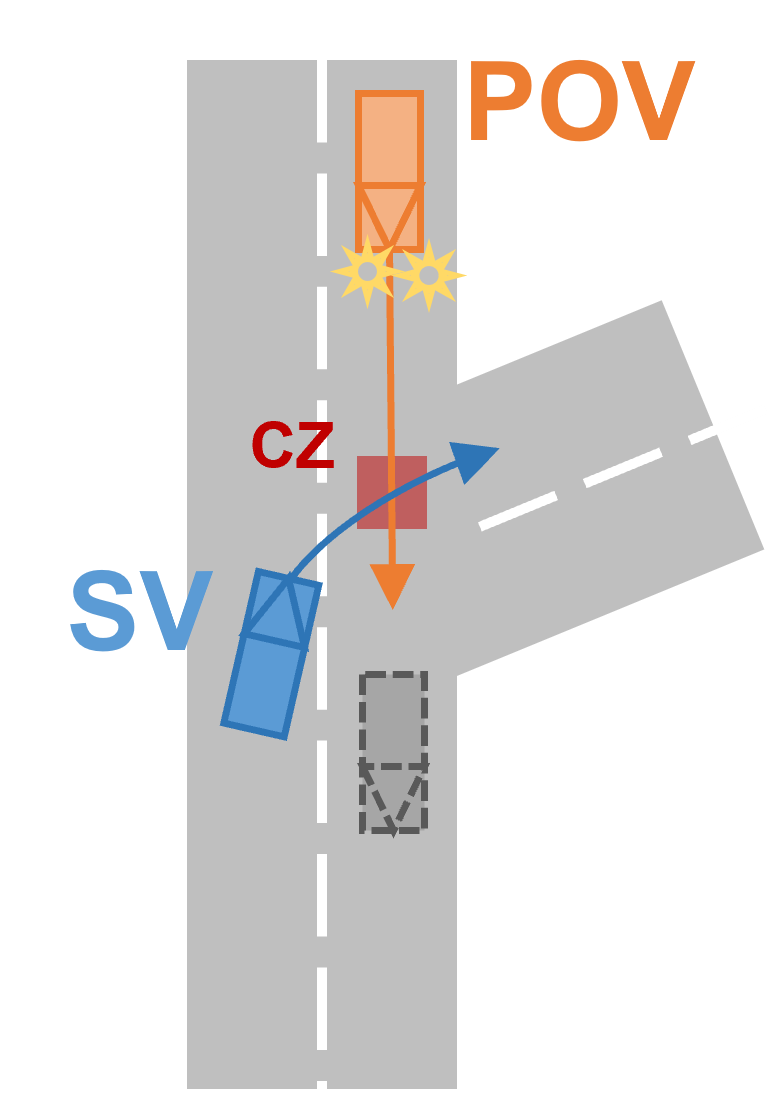}
    \label{fig:yield_pov}}
    \subfigure{
    \includegraphics[scale =0.4, width = 0.63\linewidth]{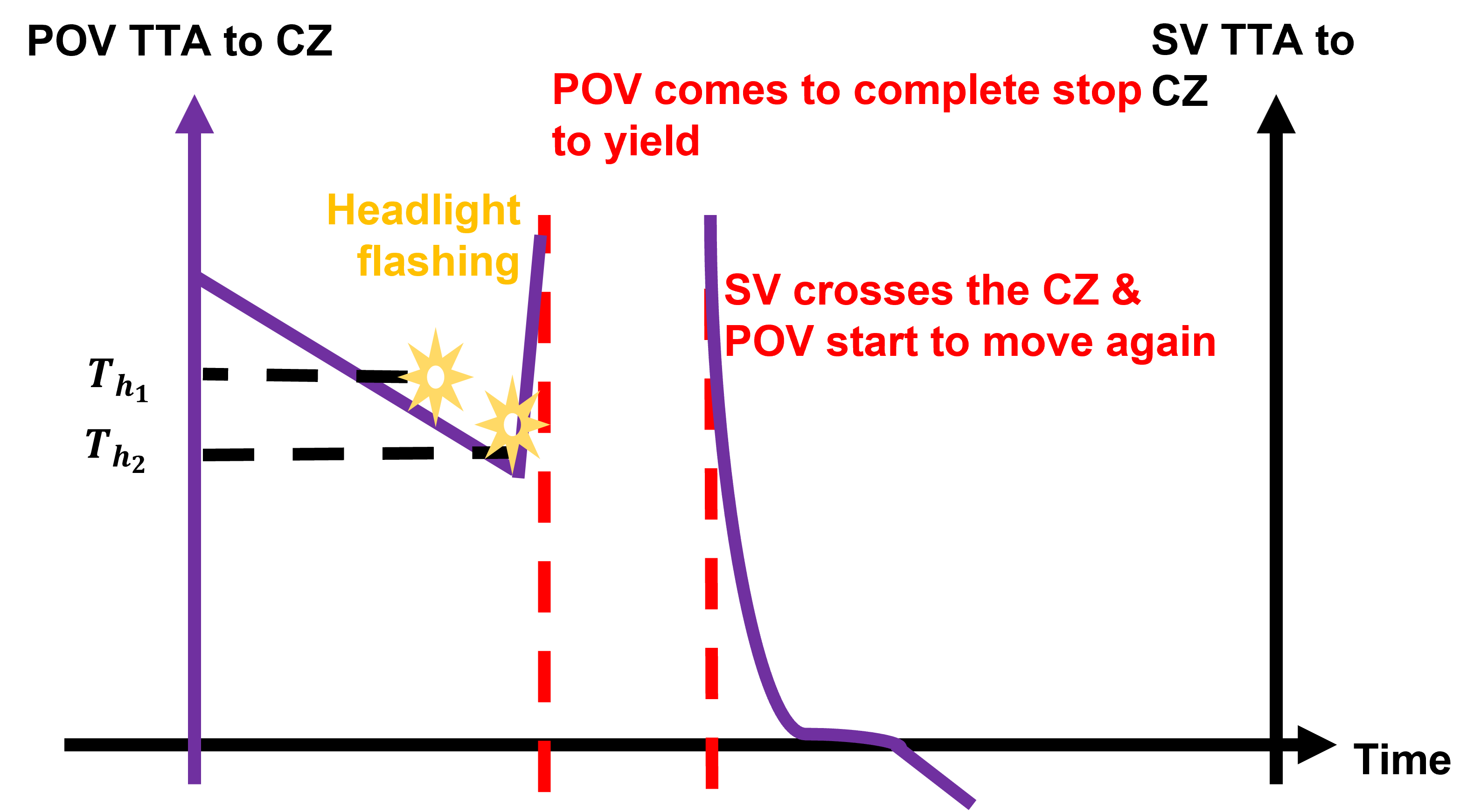}
    \label{fig:yield_pov_tta}}
    \caption{Principal other vehicle (POV) yielding to the turning subject vehicle (SV) out of courtesy (a) pictorial depiction (b) time to arrival (TTA) of the POV \& SV at the conflict zone (CZ) based schematic description. $T_{h_1}$ and $T_{h_2}$ are the headlight flashes by the POV based on SV's TTA to CZ}
    \label{fig:yield_pov_master}
    
\end{figure}

The routine driving part consisted of the SV turning against the traffic (without priority) in a T-like junction where a POV will flash headlight and give way out of courtesy to the SV. This scenario is depicted in~\figurename~\ref{fig:yield_pov}. The POV is programmed to flash the headlight twice depending on the time to arrival at the conflict zone (CZ). Here we have defined the junction where there is potential space-sharing conflict between the POV and SV as the conflict zone. The headlight flashing is a social norm followed in the UK to indicate yielding by the POV to the SV when POV has the right of way. A timeline for such a scenario is presented in~\figurename~\ref{fig:yield_pov_tta}. The POV approaches the junction at constant speed which is indicated by the decreasing time to arrival (TTA). Two consecutive headlight flashing (twice in a row for two time) happens as programmed based on POV's TTA (as presented in~\tablename~\ref{tab:headlight}) and then the POV comes to a full stop. Thus, the TTA of POV is infinite and there is a break in the TTA line. Once the SV has successfully crossed the junction the POV will start to move again. This scenario happens in everyday life where the POV can be yielding because of the traffic situation or out of altruism. 
\begin{table}[b]
    \centering
    \caption{Headlight flashing timing for the near-crash like routine yielding scenario. This is based on TTA of POV to CZ}
    \begin{tabular}{|c|c|}
    \hline
         $T_{h_1}$& $T_{h_2}$  \\
         \hline
         $2.5$ s& $2.0$ s\\
         \hline
         
    \end{tabular}
    
    \label{tab:headlight}
\end{table}

\subsubsection{SV with priority in a T-junction}\label{sec:pullout_like}
\begin{figure}
    \centering
    \subfigure{
    \includegraphics[scale =0.4, width = 0.3\linewidth]{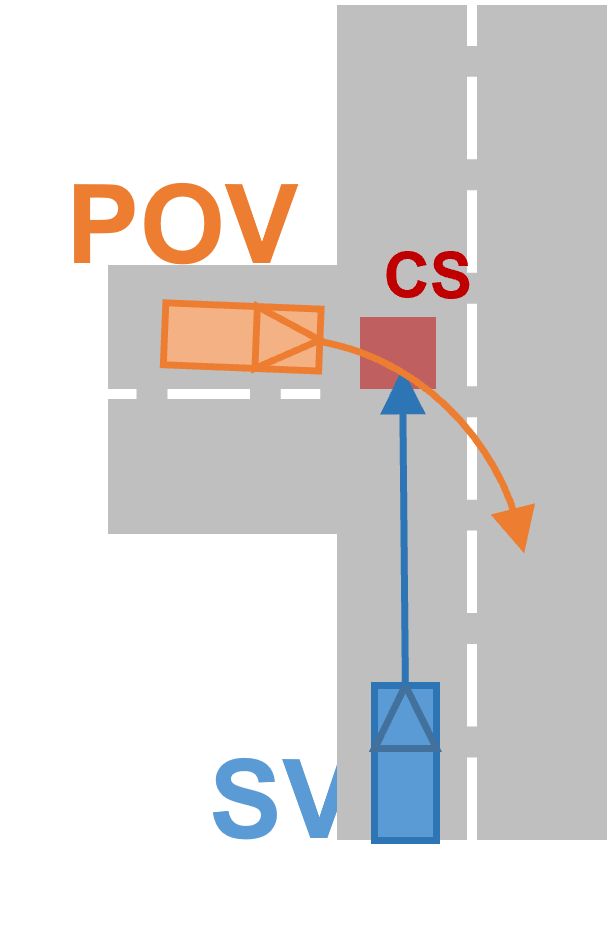}
    \label{fig:pullout_pov}}
    \subfigure{
    \includegraphics[scale =0.4, width = 0.63\linewidth]{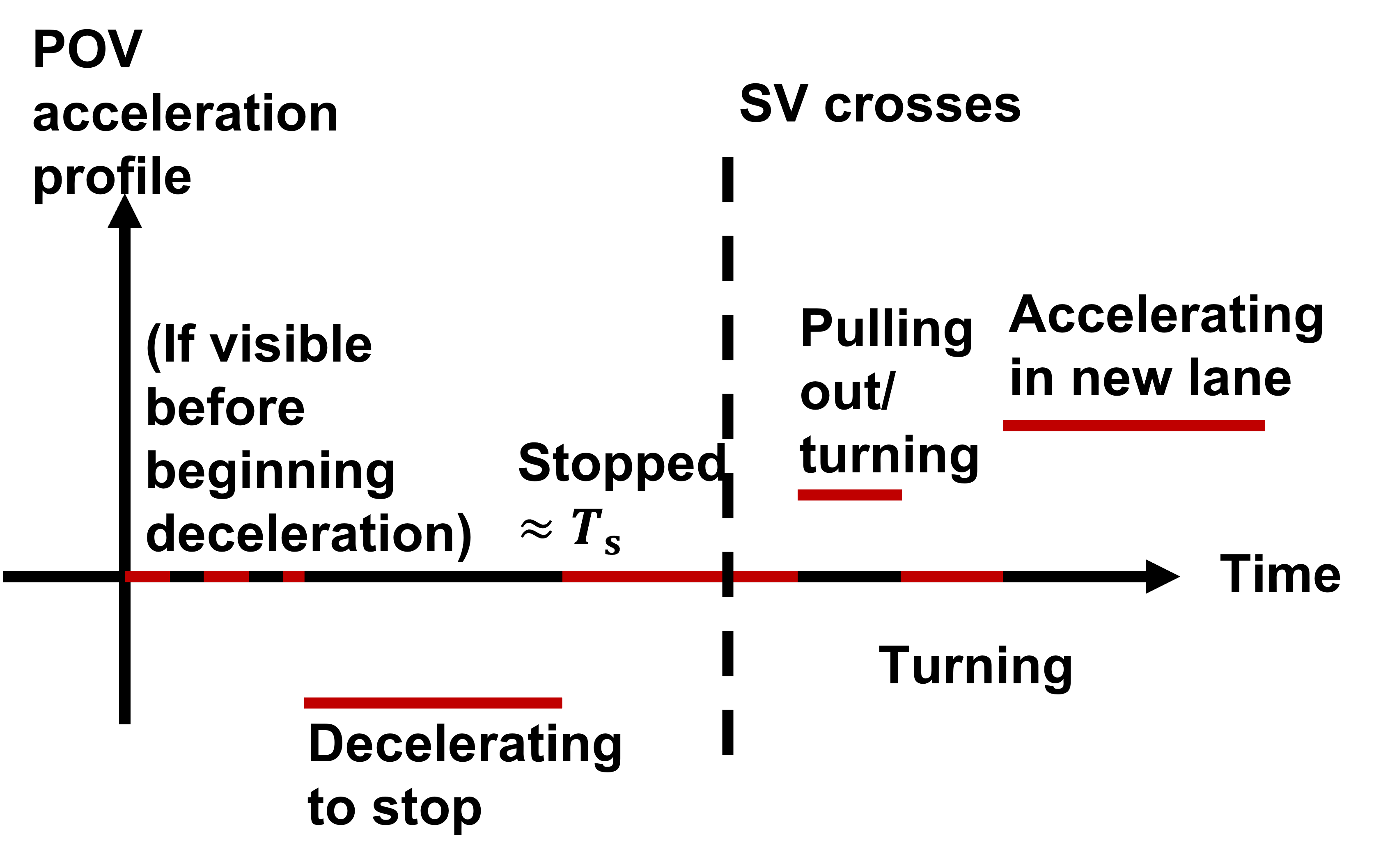}
    \label{fig:pullout_pov_tta}}
    \caption{Principal other vehicle (POV) waiting for subject vehicle (SV) to cross before pulling out and driving in opposite direction (a) pictorial depiction (b) typical acceleration profile for the POV with respect to the SV according to the scenario description}
    \label{fig:pullout_pov_master}
    
\end{figure}

Another near-crash like situation presented to the participant consisted of the SV having right of way in a straight road. A POV pulls up to a T-junction and stopped within visible range of SV as it has no priority. It turned and drove in opposite direction to the SV after SV had crossed. This situation schematic is presented in~\figurename~\ref{fig:pullout_pov_master}. All these routine scenarios where everyone follows the road rules and social norms gives the participant a typical urban driving experience. Next, we will describe our near-crash recreating scenarios. 

\subsection{Near-crash emulation} \label{sec:nc}
These are the scenarios where the SV is designed to experience a near-crash situation. It can either be an unsafe pull out by the POV in a situation similar to scenario~\ref{sec:pullout_like} or the POV not yielding after decelerating and flashing headlight to the SV turning against traffic similar to scenario~\ref{sec:like_yield}. 
\subsubsection{Unsafe pull-out by POV in a T-junction}\label{sec:unsafe_pullout}
\begin{figure}
    \centering
    \includegraphics[scale=0.4, width = 0.9\linewidth]{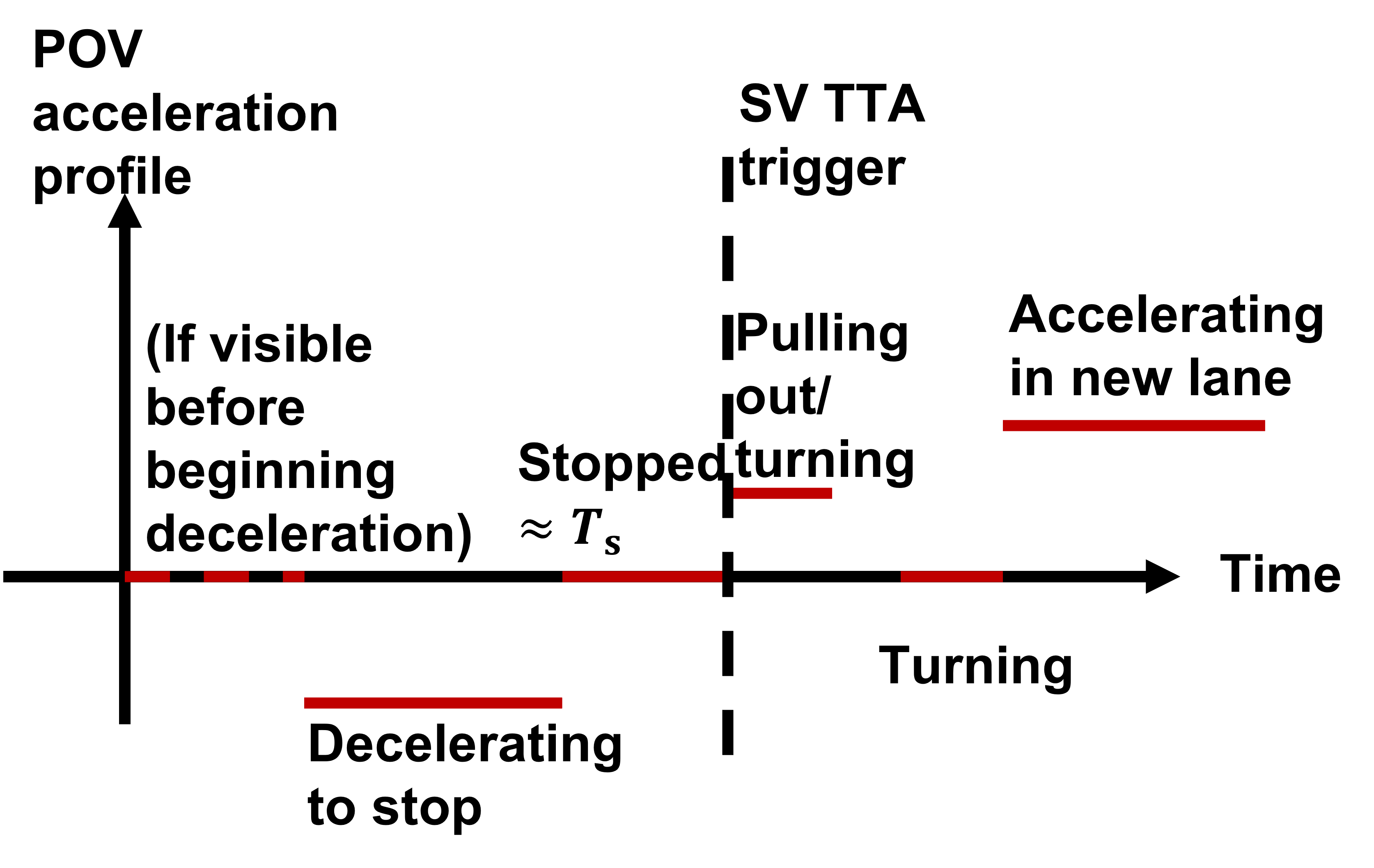}
    \caption{Acceleration profile for the POV in a near-crash pull-out by POV in a T-junction. Here we have assumed that the SV keeps a constant speed throughout the scenario. The step-like change in acceleration of POV is used to showcase the timing of the event. The POV in practice had a ramp-like acceleration profile}
    \label{fig:nc_pullout_tta}
\end{figure}

\figurename~\ref{fig:pullout_pov} shows a typical T-junction where the SV has priority and POV has to wait before crossing the SV path and joining the main roadway. Similar to some drivers who might not follow the road rules, the POV is triggered by program to cross in front of the SV in an unsafe manner. The POV will start to pull-out infront of the SV based on the TTA of the SV to the CZ. The POV will arrive at the T-junction and brake like a normal driver initially. Once, the SV has passed a pre-determined hand-tuned threshold, the POV will start to pull-out. The thresholds are presented in~\tablename~\ref{tab:nc_pullout}. The threshold was selected to ensure that the POV does an unsafe crossing. This creates a near-crash situation. We utilised two different tuning of SV's TTA trigger to recreate a more and a less critical scene. The POV's designed acceleration profile is similar to~\figurename~\ref{fig:nc_pullout_tta}.  
\begin{table}[h]
    \centering
    \caption{Unsafe pull-out scenario - SV TTA at junction based threshold}
    \begin{tabular}{|c|c|}
    \hline
         More critical & Less critical  \\
         \hline
         $3$ s& $4$ s\\
         \hline
         
    \end{tabular}
    
    \label{tab:nc_pullout}
\end{table}
\subsubsection{Mixed signalling leading to a near crash by POV with priority in a T-like junction}\label{sec:unsafe_yield}
\begin{figure}
    \centering
    \includegraphics[scale=0.4, width = 0.9\linewidth]{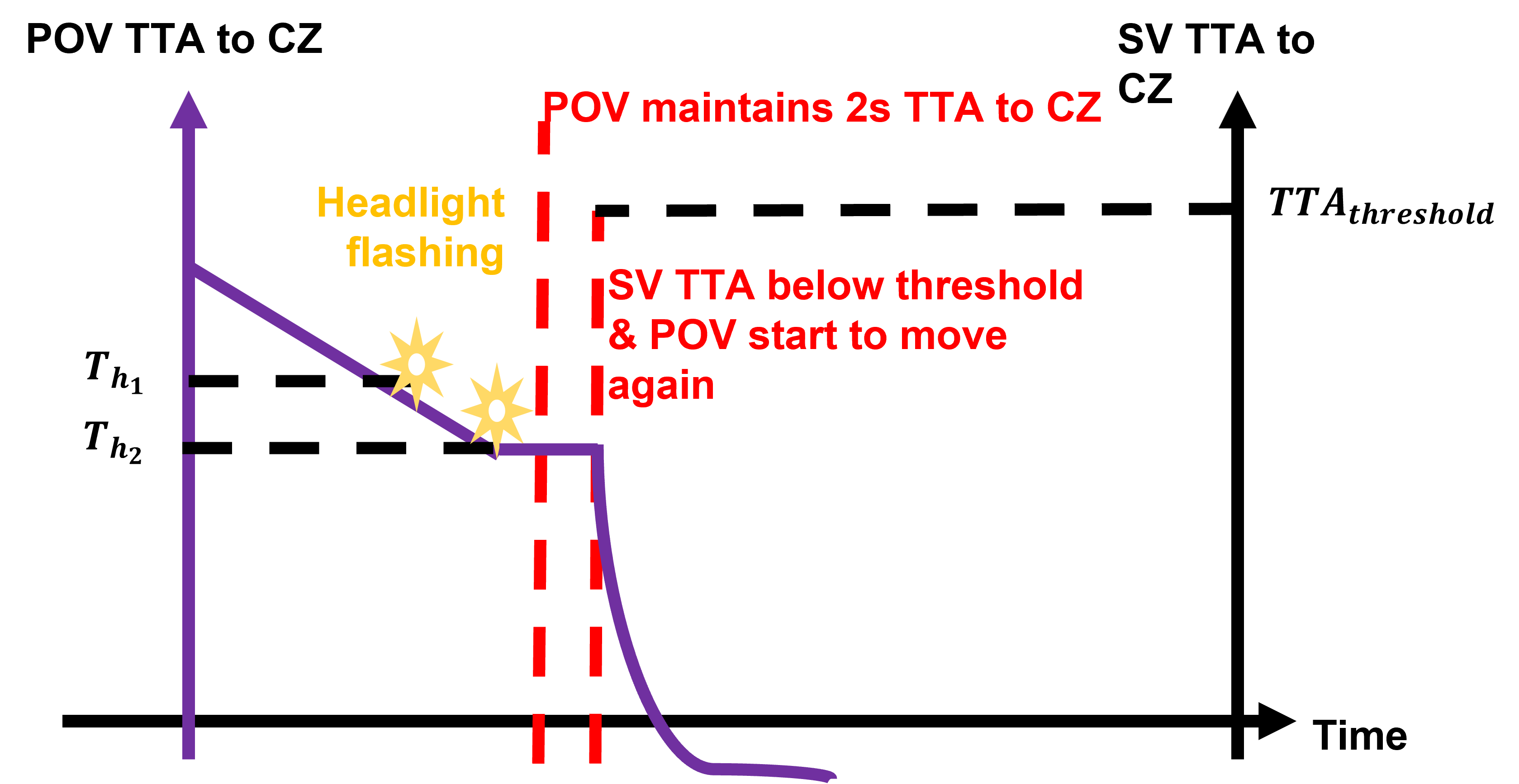}
    \caption{TTA profile for the POV in a near-crash initiated by POV in a T-like junction with headlight flashing similar to \ref{sec:like_yield}}
    \label{fig:nc_yield_tta}
\end{figure}

\figurename~\ref{fig:yield_pov} shows a POV signalling giving way to SV to cross the lane based on POV's TTA to CZ. Unlike the scenario in~\ref{sec:like_yield}, the POV will maintain the TTA after second headlight flash (decelerate as needed) and start to accelerate suddenly when the SV accepts the social signal and is within a certain TTA to the CZ (starts to turn). This again is designed with the intent to create near-crash like situation and to observe/record the participant driving behaviour. A typical timeline for the TTA change of the POV is presented in~\figurename~\ref{fig:nc_yield_tta}. Two threshold values were used to create a more critical and a less critical version of the near-crash scenario and is presented in~\tablename~\ref{tab:nc_nonyield}. The threshold values for the headlight flashes is same as the one presented in scenario~\ref{sec:like_yield} (~\tablename~\ref{tab:headlight}).
\begin{table}[h]
    \centering
    \caption{Unsafe yield scenario - SV TTA-based threshold for POV to stop yielding (decelerating) and suddenly start accelerating}
    \begin{tabular}{|c|c|}
    \hline
         More critical & Less critical  \\
         \hline
         $1.5$ s & $5$ s\\
         \hline
         
    \end{tabular}
    
    \label{tab:nc_nonyield}
\end{table}
\section{Driving simulator}

\begin{figure}
    \centering
    \includegraphics[width=0.9\linewidth]{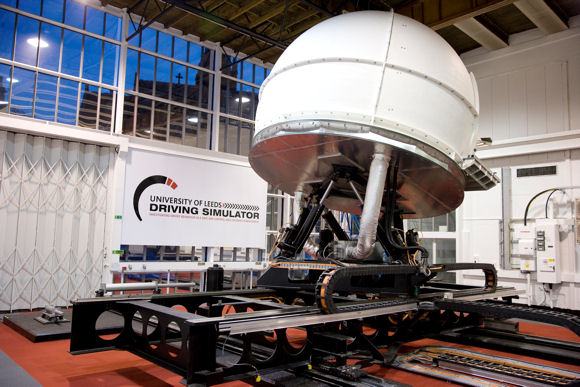}
    \caption{An artistic capture of the University of Leeds Driving Simulator (UoLDS)}
    \label{fig:UoLDS}
\end{figure}

\begin{figure}
    \centering
    \includegraphics[width=0.9\linewidth]{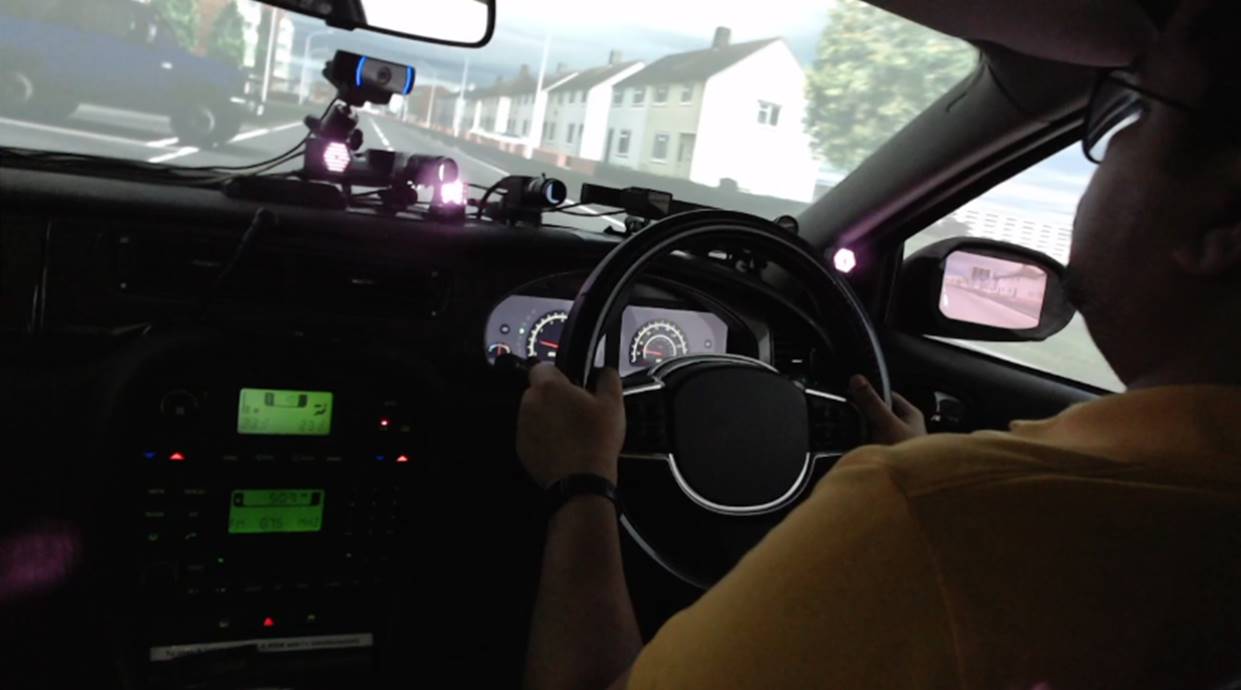}
    \caption{A wide view of the urban scene through the windshield from inside the vehicle}
    \label{fig:participant_view}
\end{figure}

We utilised the state-of-the-art driving simulator at University of Leeds (University of Leeds Driving Simulator - UoLDS). The UoLDS has a Jaguar S-type vehicle in a spherical dome. The~\figurename~\ref{fig:UoLDS} shows a digital capture of the facility. The dome is fitted with $300^o$ visual projection system. The dome itself is mounted on a 8-dof (degrees of freedom) motion platform. The $XY$ table (motion platform) provided $\pm 5$ m of translation in both longitudinal and lateral directions. We utilised the ``classical'' motion cueing algorithm from \cite{jamson2010motion}. The vehicle dynamics were simulated using a Jaguar-developed multi-body model of the XE variant. This model has been validated extensively to ensure it closely captured vehicle behaviour on a high-friction surface. In order to recreate an urban scene, the sides of the roadways were populated with appropriate buildings and landscape. A screenshot of the wide view of one such scene is showcased in~\figurename~\ref{fig:participant_view}. 
\section{Participants}
Ethics approval has been obtained from the local ethics committee prior to the start of any recruitment and data collection. Participants were recruited through emails and social media post. A total of $80$\footnote{$1$ participant could only complete one of the near-crash scenario}.The demographic summary of the participants is presented in~\tablename~\ref{tab:demographic}. We had a participants aged $20$ through $67$ with an average age of $35.95$ years and about $2$ male participants per $1$ female participant. There were more right handed participants and the average annual driving mileage was $7512$ miles. The average number of years holding a full driving license was $16.04$ years.  All effort were made to be inclusive and the recruitment was unbiased. The minimum requirements were corrected eye-vision with minimum of $3$ years driving experience in some part of the world. We had to exclude people with underlying conditions like back issues and pregnant woman to safeguard them from potential unnecessary risk and complications in case of an emergency evacuation from the driving simulator platform. 
\begin{table}
    \centering
    \caption{Demographic survey summary statistics}
    \begin{tabular}{|c|c|}
         \hline
         Average Age & $35.95$ years  \\
         % Median Age & $34$ years \\
         \hline
         No. of Males & $55$ \\
         No. of Females & $25$ \\
         \hline
         Right handed & $70$ \\
         Left handed & $10$ \\
         \hline
         Approx. annual mileage & $7512$ miles \\
         \hline
         Average No. of years holding a full UK driving license & $16.04$ years \\
         % Median No. of years holding a full UK driving license & $12$ years
         \hline
    \end{tabular}
    
    \label{tab:demographic}
\end{table}
\section{Experimental Procedure}
\begin{figure}
    \centering
    \includegraphics[ width = 0.9\linewidth]{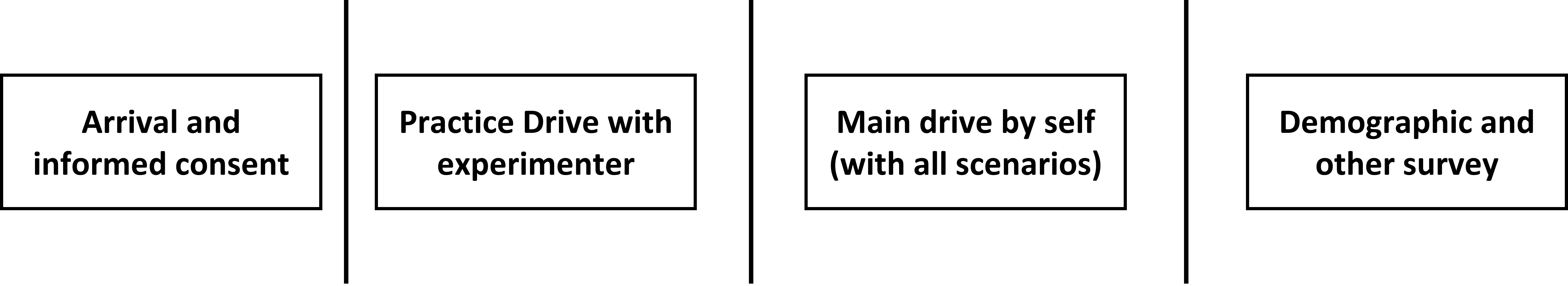}
    \label{fig:fc_data}
    \caption{A schematic for the overall data collection procedure}
    \label{fig:fc}
    
\end{figure}

 In accordance with established procedure, the participants were giving a brief introduction about the study once they arrived. They were then given a chance to provide an informed consent if they were happy to participate after considering the information shared. A typical flowchart for the data collection and overall experiment procedure is presented in~\figurename~\ref{fig:fc}.

Once the participant provided informed consent, they were briefed on the safety procedure associated with the driving simulator. Then, they were situated in the driver seat and the experimenter stayed with them for the duration of the practice drive. The practice drive was similar to the main drive with a few routine driving scenario to help the participant get accustomed to the simulator controls and the urban environment. After the end of the practice drive, the participants continued to the main drive with all the scenarios in a counter-balanced order by themselves. 
The order and the criticality (two levels of criticality) of the scenario were pseudo-randomised to ensure counter-balancing between participants. If unsafe pull out was simulated first then the other near-crash situation is presented to the participant as the last scenario and vice versa. Also, the sixteen variation of the gap acceptance scenario was counter-balanced between participants. A typical order of scenario is presented in~\tablename~\ref{tab:scenario}
\begin{table*}
    \centering
    \caption{Main drive scenario order}
    \begin{tabular}{|c|c|}
        \hline
         Scenario number & Scenario name  \\
        \hline
         1 & Yielding by POV to SV turning against traffic in a T-like junction (\ref{sec:like_yield})  \\
         \hline
         2 & Gap acceptance by SV (\ref{sec:gap_accept}) \\
         \hline
         3 & SV with priority in a T-junction (\ref{sec:pullout_like}) \\
         \hline
         4 & Traffic light turning green for SV (\ref{sec:other_sec})\\
         \hline
         5 & Near-crash scenario 1 (\ref{sec:nc}) \\
         \hline
         6 & Traffic light turning red for SV (\ref{sec:other_sec})\\
         \hline
         7 & SV turning left in a $3$ way intersection (\ref{sec:other_sec}) \\
         \hline
         8 & Gap acceptance by SV (\ref{sec:gap_accept}) \\
         \hline
         9 & near-crash scenario 2 (\ref{sec:nc}) \\
         \hline
    \end{tabular}
    
    \label{tab:scenario}
\end{table*}

A typical practice drive was approximately $10$ minutes long and the a typical main drive lasted about $25$ minutes. Participants were instructed to drive at posted speed limit of $30 mile/hour$ ($13.41 m/s$) and follow all road rules. Every participant was given the following instruction exactly: ``Please follow the route indicated by the voice prompts, and drive as you would normally, in real traffic. Assume that you want to get your destination without delay, but please follow the posted speed limit of 30 mph and respect other road users, again just as you would normally.''

 Every participant who completed at least one of the near-crash situation were asked to fill out a brief demographic survey and an open-ended questionnaire at the end of the experiment. They were also compensated for their time ($\pounds 20$) in accordance with local practice. All data were by design anonymised and no personally identifiable information was collected to ensure compliance with data protection legislation.

\section{Data description}
\href{https://osf.io/eazg5 }{Open Science Foundation - Data repository} holds the anonymised data of all participants along with metadata files\cite{srinivasan_commotions_2023}. The log data folder\footnote{2$^{nd}$ July 2024 update: We have now uploaded all data including the near-crash data.} has a metadata sub-folder to assist researchers with understanding the uploaded data. The metadata sub-folder has two files. README\_subject.txt describes the SV data file. Reference\_Data.xlsx describes the hyper-parameters and near-crash scenario state machine. Each participant data is stored as a separate sub-folder name as subx(x). The dronex(x)yy*\_runz.txt files contain the relevant data for the traffic other than the SV. The x(x) is the scenario number. Scenario number $12$ is used for all non-scenario traffic. The yy is the drone vehicle number. Scenario $7$ is always unsafe pull-out by POV in a T-junction (\ref{sec:unsafe_pullout}) and scenario $11$ is always unsafe yield by POV in a T-like junction (\ref{sec:unsafe_yield}). For the scenario $2$ and $11$, it is always the number $02$ which yields or pulls out after headlight flash. The z is for run and $1$ means practice drive and $2$ means main drive. The subjectCar*\_runz.txt file is the SV related file and the StateScenario and RunInformation provides data about the state machine and hyper-parameters used for that particular participant. 
\section{Conclusion}
We have presented a study collecting both routine and near-crash situation data in a high-fidelity driving simulator. The dataset includes 80 drivers, and a total of 160 observations of their near-crash manoeuvring across two different near-crash vehicle-vehicle interaction scenarios, each at two different levels of urgency, and 160 observations of driver gap acceptance in vehicle flows with controlled gap sequences.

\ifCLASSOPTIONcaptionsoff
 \newpage
\fi
\newpage
\bibliographystyle{IEEEtran}
\bibliography{Commotions_dataset_paper.bib}%
\end{document}